\documentclass{ws-ijmpa}
\usepackage[super,compress]{cite}
\usepackage{graphicx}
\begin{document}
\markboth{V.~M.~Mostepanenko}{Progress in constraining axion and non-Newtonian gravity
from the Casimir effect}

%
\catchline{}{}{}{}{}
%

\title{Progress in constraining axion and non-Newtonian gravity
\protect{\\} from the Casimir effect
}

\author{V.~M.~Mostepanenko
}

\address{Department of Astrophysics,
Central Astronomical Observatory at Pulkovo of the Russian Academy of Sciences,
Saint Petersburg,
196140, Russia\\
and\\
Institute of Physics, Nanotechnology and
Telecommunications, Peter the Great Saint Petersburg
Polytechnic University, Saint Petersburg, 195251, Russia\\
vmostepa@gmail.com}

\maketitle

\begin{history}
\received{24 August 2015}
\end{history}

\begin{abstract}
We consider recent progress in constraining the axion-nucleon coupling
constants and the Yukawa-type corrections to Newtonian gravity from
experiments on measuring the Casimir interaction. After a brief review
of previously obtained constraints, we concentrate on the new
Casimir-less experiment, which allows to strengthen the known results
up to  factors of 60 and 1000 for the axion-like particles and Yukawa-type
corrections, respectively. We also discuss possibilities allowing to
further strengthen the constraints on axion-nucleon coupling constants,
and propose a new experiment aiming to achieve this goal.

\end{abstract}

\ccode{PACS numbers: 14.80.Va, 12.20.Fv, 14.80.-j}

\section{Introduction}

Both axions and non-Newtonian gravity have a long history.
The Yukawa- and power-type corrections to the Newton law of gravitation
follow from the exchange of light massive or massless elementary
particles, respectively, between atoms of two macrobodies (see the detailed
review in Ref.~\citen{1}). The same corrections to Newtonian gravity
were introduced in extra-dimensional models with a low-energy
compactification scale.\cite{2,3,4,5}
The light pseudoscalar particle axion was predicted\cite{6,7}
in quantum chromodynamics as a consequence of broken Peccei-Quinn
symmetry.\cite{8} The latter was postulated in order to avoid strong
PC violation and large electric dipole moment of a neutron which are
excluded by the experimental data. Axions play an important role in
astrophysics and cosmology because they are considered as possible
constituents of dark matter.\cite{9,10}

Strong constraints on the corrections to Newton gravitational law
have been obtained from the astronomical observations, geophysical
experiments and from the experiments of E\"{o}tvos and Cavendish
type.\cite{1} In the beginning of the 21st century several new
E\"{o}tvos- and Cavendish-type experiments were performed in order
to strengthen these constraints in the interaction region below
several hundreds micrometers (see Ref.~\citen{11} for a review).
However, no signature of deviations from the Newtonian gravity
have been observed.

During the last few decades a lot of experiments on the search of
axions were performed using their interactions with photons,
electrons and nucleons (see Refs.~\citen{9} and
\citen{12,13,14} for a review).  In addition to the laboratory
experiments, many constraints on the axion parameters were
obtained from astrophysics and cosmology, e.g., from stellar
cooling, neutrino data of supernova SN\,\,1987A, and attempts
to observe axions emitted by the Sun or belonging to the dark
matter.\cite{13,14,15} As a result, the originally introduced
axions were constrained to a very narrow band in the parameter
space\cite{15} and many different models of the so-called
axion-like particles were proposed. The most of these models
can be attributed to one of the two sets of models of QCD
(or hadronic) axions\cite{16,17} or Grand Unified Theory (GUT)
axions.\cite{18,19}

It has long been known that strong constraints on the
Yukawa-type\cite{20} and power-type\cite{21} corrections to
Newtonian gravitational law can be obtained from measurements
of the van der Waals and Casimir force. Precise measurements
of  the Casimir interaction performed during the last few
years\cite{22,23} allowed for significant strengthening of
previously known constraints on the Yukawa-type corrections
to Newton's law in the interaction range below one micrometer
(see Refs.~\citen{24} and \citen{25} for a review).
Recently it was shown\cite{26,27,28,29} that the same
measurements place strong constraints  on the coupling
constants of the GUT axions to nucleons in the wide range of
axion masses (see Ref.~\citen{30} for a review).

In this paper we describe the most modern constraints on the
axion-nucleon coupling constants and Yukawa-type corrections to
Newton's law obtained from the Casimir effect. In Sec.~2 we
consider the effective potentials following from different
elementary processes and explain what kind of constraints could
be obtained from each of them. We also attract reader's
attention to one unresolved problem. In Sec.~3 we briefly
review the constraints on the coupling constants of the GUT
axions to nucleons following from measurements of the thermal
Casimir-Polder force. In Sec.~4 the same is done regarding the
constraints following from measurements of the gradient of the
Casimir force, Casimir pressure and lateral Casimir force.
Sections 5 and 6 are devoted to the most strong constraints
on the Yukawa-type corrections to Newton's law and axion-nucleon
coupling constants obtained from the new Casimir-less
experiment. Section 7 contains our conclusions and discussion
including the proposal of a pioneer Casimir experiment
exploiting the polarized test bodies.

We use the system of units in which $\hbar=c=1$.

\section{Effective Potentials}

We begin with an exchange of one light scalar particle of mass $M$
between two particles with masses $m_1$ and $m_2$ located
at the points
 $\mbox{\boldmath$r$}_1$ and $\mbox{\boldmath$r$}_2$.
It is common knowledge that this process results in the
spin-independent Yukawa-type effective potential.\cite{1}
It is conventional to parametrize it as a correction to
Newton's law
\begin{equation}
V(|\mbox{\boldmath$r$}_{12}|)=-
\frac{Gm_1m_2}{|\mbox{\boldmath$r$}_{12}|}\,\left(1+
\alpha e^{-|\mbox{\boldmath$r$}_{12}|/\lambda}\right),
\label{eq1}
\end{equation}
\noindent
where
$\mbox{\boldmath$r$}_{12}=\mbox{\boldmath$r$}_1-\mbox{\boldmath$r$}_2$,
$G$ is the gravitational constant.
$\alpha$ is a dimensionless interaction constant, and $\lambda=1/M$
 is the Compton wavelength of light scalar particle (i.e., the interaction range).
If the Yukawa-type correction originates from extra-dimensional models
(see Sec.~1), the quantity $\lambda$ characterizes the size of a compact manifold
formed by the extra dimensions.\cite{2,3}

If the scalar particle is massless, $\lambda\to\infty$ and we return from
(\ref{eq1}) to the potential which is inverse proportional to separation
similar to the Newtonian potential.
There are also power-type effective potentials with higher powers which are
usually parametrized
as corrections to Newtonian  gravity
\begin{equation}
V_n(|\mbox{\boldmath$r$}_{12}|)=-
\frac{Gm_1m_2}{|\mbox{\boldmath$r$}_{12}|}\,\left[1+
\Lambda_n
\left(\frac{r_0}{|\mbox{\boldmath$r$}_{12}|}\right)^{n-1}
\right],
\label{eq2}
\end{equation}
\noindent
where $n=1,\,2,\,3,\,\ldots\,$, $\Lambda_n$ is a dimensionless interaction
constant,  and $r_0=10^{-15}\,$m is chosen to preserve the correct
dimension of energy at different $n$.
The effective potentials of power type arise due to an exchange of an
even number of massless pseudoscalar particles, such as arions,\cite{31}
and also due to exchanges of two neutrinos, two goldstinos, or other
massless fermions.\cite{32}

The Casimir effect was used to constrain the axion-nucleon coupling constants.
Because of this, here we consider the effective potentials arising due to
the exchange of axions between protons and neutrons belonging to two
different test bodies.
The interaction of GUT axion-like particles $a$  with nucleons $\psi$
is described by the pseudoscalar Lagrangian density\cite{9,13,33}
\begin{equation}
{\cal L}_{\rm ps}=-ig_{ak}\bar{\psi}\gamma_5\psi a,
\label{eq3}
\end{equation}
\noindent
where $g_{ak}$ is the dimensionless coupling constant of an axion to
 a proton ($k=p$) or a neutron ($k=n$), and $\gamma_n$ with
 $n=0,\,1,\,2,\,3,\,5$ are the Dirac matrices.
The interaction of QCD  axions (which are pseudo-Namby-Goldstone bosons)
with nucleons is described by the pseudovector Lagrangian density\cite{9,13}
\begin{equation}
{\cal L}_{\rm pv}=\frac{g_{ak}}{2m_a}\bar{\psi}\gamma_5\gamma_{\mu}\psi
\partial^{\mu}a,
\label{eq4}
\end{equation}
\noindent
where  $m_a$ is the axion mass and
the effective interaction constant $g_{ak}/(2m_a)$ is dimensional.
Note that the mass and interaction constant of  the GUT axions  are the
independent parameters, whereas
the mass and interaction constant of  the QCD axions  are connected
by some relationship.\cite{9}

It is interesting that on a tree level both Lagrangian densities
(\ref{eq3}) and (\ref{eq4}) lead to one and the same action after an
integration by parts. As a result, both (\ref{eq3}) and (\ref{eq4}) lead to
common effective potential due to the exchange of either one GUT or one
QCD axion between two nucleons\cite{34,35}
\begin{eqnarray}
&&
V_{kl}(\mbox{\boldmath$r$}_{12};\mbox{\boldmath$\sigma$}_{1},\mbox{\boldmath$\sigma$}_{2})
=\frac{g_{ak}g_{al}}{16\pi m_km_l}\left[
\vphantom{\left(
\frac{m_a}{|\mbox{\boldmath$r$}_{12}|^2}
\frac{1}{|\mbox{\boldmath$r$}_{12}|^3}\right)}
(\mbox{\boldmath$\sigma$}_{1}\cdot\mbox{\boldmath$n$})
(\mbox{\boldmath$\sigma$}_{2}\cdot\mbox{\boldmath$n$})\right.
\left(
\frac{m_a^2}{|\mbox{\boldmath$r$}_{12}|}+
\frac{3m_a}{|\mbox{\boldmath$r$}_{12}|^2}+
\frac{3}{|\mbox{\boldmath$r$}_{12}|^3}\right)
\nonumber\\
&&~~~~
\left.-
(\mbox{\boldmath$\sigma$}_{1}\cdot\mbox{\boldmath$\sigma$}_{2})
\left(
\frac{m_a}{|\mbox{\boldmath$r$}_{12}|^2}+
\frac{1}{|\mbox{\boldmath$r$}_{12}|^3}\right)
\right]\,e^{-m_a|\mbox{\boldmath$r$}_{12}|},
\label{eq5}
\end{eqnarray}
\noindent
where
$\mbox{\boldmath$n$}=\mbox{\boldmath$r$}_{12}/|\mbox{\boldmath$r$}_{12}|$
is the unit vector, and  $m_k,\,m_l$ and
$\mbox{\boldmath$\sigma$}_{1},\,\mbox{\boldmath$\sigma$}_{2}$
are the nucleon masses and spins, respectively.
This potential, however,  depends on the nucleon spins.
The resulting interaction between two unpolarized test bodies, used in
already performed experiments on measuring the Casimir force,\cite{22,23}
 averages to zero.
Therefore, these experiments are not suitable for constraining the
axion-nucleon interaction caused by the exchange of one axion.

The spin-independent effective potential is obtained from the exchange
of two GUT axions between two nucleons. The respective calculation is
straightforward because the quantum field theory with the Lagrangian
density (\ref{eq3}) is renormalizable. The result is\cite{35,36,37}
\begin{equation}
V_{kl}(|\mbox{\boldmath$r$}_{12}|)=
-\frac{g_{ak}^2g_{al}^2}{32\pi^3m_km_l}\,
\frac{m_a}{|\mbox{\boldmath$r$}_{12}|^2}
K_1(2m_a|\mbox{\boldmath$r$}_{12}|),
\label{eq6}
\end{equation}
\noindent
where $K_1(z)$ is the modified Bessel function of the second kind,
and it is assumed that $|\mbox{\boldmath$r$}_{12}|\gg 1/m_{k,l}$.
In the limiting case $m_a\to 0$ one obtains from (\ref{eq6}) an
attractive power-type potential
\begin{equation}
V_{kl}(|\mbox{\boldmath$r$}_{12}|)=
-\frac{g_{ak}^2g_{al}^2}{64\pi^3m_km_l}\,
\frac{1}{|\mbox{\boldmath$r$}_{12}|^3},
\label{eq7}
\end{equation}
\noindent
i.e., the same result as follows from the exchange of two massless
pseudoscalar particles between two fermions.\cite{31}

The exchange of two QCD axions between two nucleons is a much more
complicated process. The point is that
the quantum field theory with the Lagrangian
density (\ref{eq4}) is not renormalizable.
According to our knowledge, the effective potential due to exchange
of two massive QCD axions between two fermions is not yet obtained.
As to the case of two massless pseudoscalar particles, the Lagrangian
density (\ref{eq4}) leads to\cite{38}
\begin{equation}
V_{kl}(|\mbox{\boldmath$r$}_{12}|)=
\frac{3g_{ak}^2g_{al}^2}{128\pi^3m_k^2m_l^2}\,
\frac{1}{|\mbox{\boldmath$r$}_{12}|^5},
\label{eq8}
\end{equation}
\noindent
which is, surprisingly, a repulsive potential, as opposed to the
effective potentials (\ref{eq6}) and (\ref{eq7}) which are the attractive ones.

This problem invites further investigation.

\section{Constraints on Axion-Like Particles from Measuring
the Casimir-Polder Force}

In this and next sections we briefly review the constraints on axion-nucleon
coupling constants obtained from several laboratory experiments.
This is done under the natural assumption\cite{35} that $g_{an}=g_{ap}$.
Before dealing with the Casimir-Polder force, we mention the constraints
following from the magnetometer measurements and several gravitational experiments.

\begin{figure}[b]
\vspace*{-8.5cm}
\hspace*{1cm}
\centerline{\includegraphics[width=18.8cm]{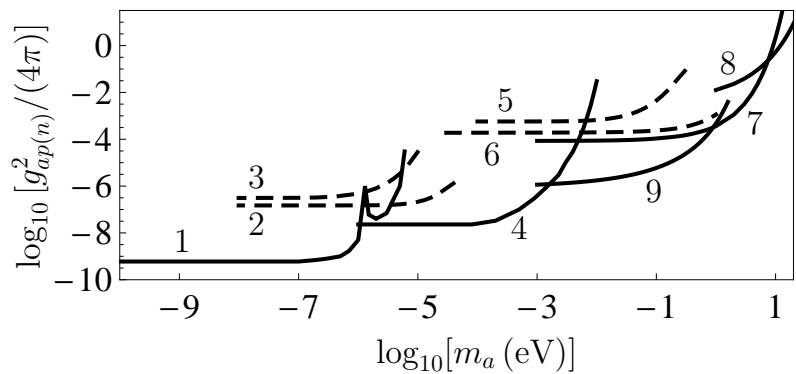}}
\vspace*{-13.5cm}
\caption{Laboratory constraints on the axion-nucleon coupling constants from
the magnetometer measurements (line 1), Cavendish-type (lines 2, 4) and
E\"{o}tvos-type (line 3) experiments, from measurements of the Casimir-Polder
force (line 5), of the gradient of the Casimir force (line 6), of the Casimir
pressure (line 7), of the lateral Casimir force between corrugated
surfaces (line 8), and from the Casimir-less experiment (line 9) are shown
as functions of the axion mass. The regions above each line are prohibited,
and below each line are allowed.}
\end{figure}
The magnetometer using spin-polarized K and ${}^3$He atoms was used\cite{39}
to obtain the constraints on $g_{an}$ and $m_a$ shown by the line 1 in Fig.~1.
Note that for all lines in Fig.~1 the axion mass and the axion-nucleon
coupling constant are considered as independent parameters.
The regions below the lines are allowed and the regions above the lines are
prohibited by the results or respective experiment.

The gravitational experiments of E\"{o}tvos and Cavendish types have long
been used for constraining the axion-nucleon coupling constants for the GUT
axions. In doing so, the additional force due to two-axion exchange was
calculated by the additive summation of internucleonic potentials (\ref{eq6})
over the volumes of the test bodies taking into account their isotopic
composition.\cite{1,35} The same method was later applied to experiments on
measuring the Casimir interaction.\cite{26,27,28,29}
The dashed lines 2 and 3 in Fig.~1 show the constraints obtained\cite{35}
from the Cavendish-type experiments\cite{40,41} and from the
E\"{o}tvos-type experiment,\cite{42} respectively.
The solid line 4 indicates the most modern gravitational constraints on
the GUT axion obtained\cite{43}
from the recent Cavendish-type experiment.\cite{44}
As is seen in Fig.~1, the line 4 provides rather strong constraints on the
masses and coupling constants of GUT axions in the mass interval centered
at $m_a=10\,\mu$eV. With increasing axion mass, the strength of these
constraints quickly decreases.

Now we consider the constraints on $g_{an}$ and $m_a$ of the GUT axions
obtained\cite{26} from measurements of the thermal Casimir-Polder force
between ${}^{87}$Rb atoms belonging to a Bose-Einstein condensate cloud
and an amorphous SiO${}_2$ plate.\cite{45} In this dynamic experiment,
the condensate cloud was placed in a magnetic trap near a plate, and the
dipole oscillations in it with the natural frequency $\omega_{0z}$ and
some constant amplitude were excited in the direction $z$ perpendicular
to the plate. The distance $d$ between the center of mass of a cloud and
a plate varied from 6.88 to $11\,\mu$m, i.e., in the thermal regime.
The Casimir-Polder force, acting  between ${}^{87}$Rb atoms and a plate,
leads to a shift of the oscillation frequency $\omega_{0z}$ to some value
$\omega_z$, and the relative frequency shift
\begin{equation}
\gamma_z=\frac{|\omega_{0z}-\omega_z|}{\omega_{0z}}
\label{eq9}
\end{equation}
\noindent
was measured as a function of separation $d$.
By solving the oscillator problem under the influence of external
(Casimir-Polder) force, the quantity $\gamma_z$ was also calculated.
For this purpose, the Casimir-Polder force  between ${}^{87}$Rb atoms
and a plate was calculated using the Lifshitz theory of dispersion
forces\cite{23,46} with subsequent averaging over the condensate
cloud.

The comparison of experiment with theory has led to a conclusion that
the calculation results are in a good agreement with the data in the
limits of the experimental errors $\Delta\gamma_z(d)$ determined at the
67\% confidence level.\cite{45} This conclusion was obtained by omitting
the role of free charge carriers in the SiO${}_2$ plate which are in fact
present at any nonzero temperature. As was shown later,\cite{47} the
account of charge carriers in the plate material results in an exclusion
of the theoretical predictions of the Lifshitz theory by the data
(see the discussion in Refs.~\citen{22} and \citen{23}).

The two-axion exchange between protons and neutrons belonging to
${}^{87}$Rb atoms and to SiO${}_2$ plate results in some additional force.
The respective additional frequency shift $\gamma_z^{\rm add}$ was
calculated by the additive summation of potentials (\ref{eq6}) with
subsequent averaging over the condensate cloud.\cite{26}
This additional shift was not observed and, thus, it should be
constrained in the limits of the experimental error
\begin{equation}
\gamma_z^{\rm add}(d)\leq\Delta\gamma_z(d).
\label{eq10}
\end{equation}

The constraints on the axion-nucleon coupling constant following from Eq.~(\ref{eq10})
are shown by the dashed line 5 in Fig.~1. As is seen in the figure, the strength of these
constraints decreases with increasing axion mass, but it is stronger than that of
line 4 in the region of axion masses centered at $m_a=10^{-2}\,$eV.

Several other experiments on measuring the Casimir interaction used for
constraining the parameters of the GUT axions are considered in the next section.

\section{Constraints on Axion-Like Particles from
Various Measurements of the Casimir Interaction}

The most precise measurements of the Casimir interaction were performed by means
of an atomic force microscope and micromechanical oscillator in the configuration
of a metal-coated sphere and a metal-coated plate.\cite{22,23}
Here, we concentrate on the case of Au coatings and do not consider other metals,
such as Al, Cu or Ni, because they result in weaker constraints.

\subsection{Gradient of the Casimir force}

The gradient of the Casimir force between Au-coated surfaces of a hollow glass
sphere and a sapphire plate was measured\cite{48,49} by means of dynamic atomic
force microscope over the separation region from 235 to 500\,nm.
Below the Au-coating both test bodies were covered with various additional
material layers. This does not influence the Casimir force but should be taken
into account in calculation of the additional force due to two-axion exchange.
The measured gradients $F_C^{\prime}(d)$ were compared with theoretical
predictions of the Lifshitz theory and found to be in a good agreement under
a condition that the relaxation properties of conduction electrons are omitted
in computations.\cite{48} Note that the measurement data of all precise experiments
on the Casimir force agree with theory only when
the relaxation properties of conduction electrons are omitted (for metals) or the
free charge carriers are disregarded (for dielectrics; see Sec.~3).
The two experiments which disagree with this observation are shown to be
erroneous.\cite{50,51,52}

The gradient of the additional force $F_{\rm add}^{\prime}(d)$ due to two-axion
exchange acting between a sphere and a plate was calculated\cite{27} using
Eq.~(\ref{eq6}) with account of the layer structure of these test bodies and
their isotopic composition.\cite{1}
In the limits of the measurement error $\Delta F_C^{\prime}(d)$ (which is used
here at the 67\% confidence level) the theoretical predictions for the gradient
of the Casimir force have been confirmed. Thus, one arrives at
\begin{equation}
F_{\rm add}^{\prime}(d)\leq\Delta F_C^{\prime}(d).
\label{eq11}
\end{equation}
\noindent
The constraints on the axion-nucleon coupling constants following from (\ref{eq11})
are presented by the dashed line 6 in Fig.~1 as the function of an axion mass.
The comparison of the lines 6 and 5 shows that the constraints obtained from
measurements of the gradient of the Casimir force are stronger than those obtained from
measurements of the Casimir-Polder force up to a factor of 170.

\subsection{Casimir pressure}

The Casimir pressure between two Au-coated parallel plates was determined
dynamically by means of micromechanical oscillator in the configuration of a sapphire
sphere and a silicon plate.\cite{53,54} Both the test bodies were coated by the
layers of Cr prior to a Au coating. The indirectly measured Casimir pressure was
compared with the Lifshitz theory and found to be in a very good agreement in the
limits of the experimental errors $\Delta P_C(d)$. The latter were determined at a
95\% confidence level, but recalculated to a 67\%  confidence level in order to
obtain the constraints comparable with those discussed above. The comparison was
made with omitted relaxation properties of conduction electrons.

The additional pressure $P_{\rm add}(d)$ due to two-axion exchange between nucleons
of the test bodies taking into account their layer structure was calculated (see
Ref.~\citen{28} for details). The constraints on the axion-nucleon coupling constants
and masses were obtained from the inequality
\begin{equation}
|P_{\rm add}(d)|\leq\Delta P_C(d).
\label{eq12}
\end{equation}
\noindent
In Fig.~1 these constraints are shown by the solid line 7. As is seen in the figure,
they significantly improve the constraints of lines 5 and 6. Specifically, at
$m_a=1\,$eV the constraints of line 7 are by a factor of 3.2 stronger than
those of line 6.

\subsection{Lateral Casimir force}

It has been known that if the test bodies are corrugated and there is some phase
shift $\varphi\neq 0$ between corrugations, then the Casimir force acts not only
perpendicular to their surfaces, but also along them.\cite{22,23} This is the
so-called lateral Casimir force. The lateral Casimir force between two Au-coated
surfaces of a sphere and a plate covered with the longitudinal sinusoidal
corrugations was first observed in Refs.~\citen{55} and \citen{56} and found
to be in qualitative agreement with theory based on the proximity force
approximation. The precise measurements of the lateral Casimir force in similar
configuration (but with shorter corrugation period equal to $\Lambda=574.4\,$nm),
as a function of the phase shift, were performed over the separation region from
120 to 190\,nm between the mean levels of corrugations.\cite{57,58}
The lateral Casimir force was independently computed using the exact theory,
which generalizes the Lifshitz theory for the case of arbitrary shaped
boundary surfaces of the test bodies. The comparison between experiment and
theory demonstrated a good agreement in the limits of the experimental error
of force measurements $\Delta F_C^{\,\rm lat}(d)$ determined at a 95\%
confidence level.\cite{57,58}

In Ref.~\citen{29} the effective potential (\ref{eq6}) was used to calculate the
additional force, $F_{\rm add}^{\,\rm lat}(d)$, which arises between the
corrugated test bodies due to two-axion exchange. This was done taking into
account the material properties of a plate (a hard epoxy coated with a layer of Au)
and a sphere (a polystyrene core coated with a layer of Cr and then with a layer
of Au). The maximum magnitude of the additional force at fixed separation is
achieved at the phase shift $\varphi_0=\pi/2$. Taking into account that no
deviations were observed from theoretical predictions, the constraints on the
parameters of GUT axions were found from the inequality
\begin{equation}
\max|F_{\rm add}^{\,\rm lat}(d)|\leq\Delta F_C^{\,\rm lat}(d),
\label{eq13}
\end{equation}
\noindent
where the experimental error was recalculated to the 67\% confidence level.
The obtained constraints are shown by the solid line 8 in Fig.~1 in the mass
range from 1 to 20\,eV. For $m_a\geq 8\,$eV they are stronger that the constraints
obtained by means of micromechanical oscillator (the line 7 in Fig.~1).

Note that the normal Casimir force between sinusoidally corrugated bodies, acting
perpendicular to the surfaces, has also been measured.\cite{59,60}
This experiment does not lead to stronger constraints on an axion, but is used in the
next section when discussing the Yukawa-type corrections to Newtonian gravity.

\section{New Casimir-Less Experiment and Stronger Constraints on
Non-Newtonian Gravity}

First we briefly list the strongest constraints on the parameters $\alpha$ and $\lambda$
of the effective potential (\ref{eq1}) obtained from measurements of the Casimir
interaction [the strongest present constraints on the constants of power-type
potentials (\ref{eq2}) follow from the E\"{o}tvos- and Cavendish-type
experiments\cite{23,25}].

The Yukawa-type interaction energy between the test bodies in experiments on
measuring the Casimir interaction is obtained by the integration of the effective
potential (\ref{eq4}) over the volumes of the text bodies. In doing so,
the Newtonian contribution to the results at submicrometer separations turns out
to be negligibly small, as compared to the errors in force measurements, and
can be neglected. The most strong constraints on $\alpha$ and $\lambda$ in the
micrometer and submicrometer interaction range are summarized in Fig.~2.
The line 1 was obtained\cite{24} from measurements of the lateral Casimir force
between corrugated surfaces.\cite{57,58}
This experiment was discussed in the previous section. It leads to the strongest
constraints in the interaction region from $\lambda=1.6$ to 11.6\,nm.
The constraints on $\alpha,\,\lambda$ found\cite{61} from measurements of the
normal Casimir force between corrugated surfaces at the angle of 2.4\,${}^{\circ}$
between corrugations,\cite{59,60} are shown by the line 2. These constraints are
the strongest in the interaction region from 11.6 to 17.2\,nm.
The line 3 indicates the constraints found\cite{53,54} from indirect measurement
of the Casimir pressure by means of micromechanical oscillator (see Sec.~4.2).
Until very recently they were the strongest ones over the wider interaction region
from 17.2 to 89\,nm.
\begin{figure}[b]
\vspace*{-7.2cm}
\hspace*{1cm}
\centerline{\includegraphics[width=18.8cm]{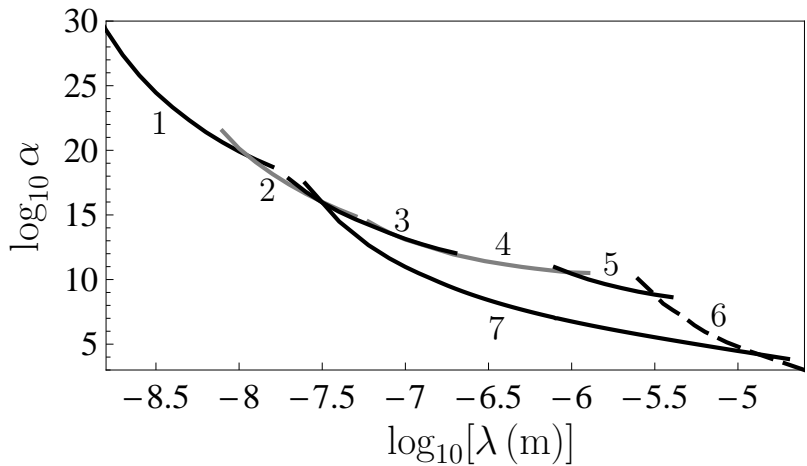}}
\vspace*{-13.8cm}
\caption{Constraints on the Yukawa-type corrections to Newton's law
of gravitation obtained from measurements
of the lateral and normal Casimir force between corrugated surfaces
(line 1 and 2, respectively),
of the Casimir pressure (line 3),
from the old Casimir-less experiment (line 4),
from measurements of the Casimir force between a plate and a spherical lens
(line 5),
from the Cavendish-type experiment (line 6),  and
from the new Casimir-less experiment (line 7) are shown
as functions of the interaction range. The regions above each line are prohibited,
and below each line are allowed.}
\end{figure}

At larger $\lambda$ strong constraints on $\alpha$ have been obtained from
measurements of another type, the so-called Casimir-less experiment.\cite{62}
In this approach the Casimir force was nullified by using the difference force
measurement, and the Yukawa-type force was restricted by the force sensitivity
of the measurement device (see below for the improved version of such kind
experiment). The constraint found in this way\cite{62} are shown by the line 4
in Fig.~2. They extend from $\lambda=89\,$nm to $\lambda=891\,$nm.
Finally, the line 5 demonstrates the constraints obtained\cite{63} from
measurements of the Casimir force between Au-coated surfaces of a plate and
a spherical lens of centimeter-size radius. These constraints are the most
strong in the interval $0.891\,\mu\mbox{m}<\lambda<3.16\,\mu$m.
At $\lambda\geq 3.16\,\mu$m the strongest constraints on $\alpha$ shown
by the dashed line 6 follow from the Cavendish-type experiments.\cite{44,64,65}

Very recently, new and significantly improved Casimir-less experiment has been
performed.\cite{66} In this experiment, a differential force between a Au-coated
sphere and either a Au sector or a Si sector of the structured disc deposited
on a Si substrate and coated with overlayers of Cr and Au was measured.
In such  away, the contribution of the Casimir force between two Au surfaces
to the differential force was nullified. Then the measurement result was
determined solely by the difference in the forces due to the exchange of some
hypothetical particles when the sphere is above  Au or Si sectors.
The differential Yukawa-type force
$F_{\rm add,\,Yu}^{\rm Au,\,Au}(d)-F_{\rm add,\,Yu}^{\rm Au,\,Si}(d)$
was calculated with account of material composition of the test bodies.\cite{66}
Taking into account that no statistically meaningful signal was observed,
the constraints on $\alpha$ and $\lambda$ were obtained from
an inequality\cite{66}
\begin{equation}
|F_{\rm add,\,Yu}^{\rm Au,\,Au}(d)-F_{\rm add,\,Yu}^{\rm Au,\,Si}(d)|
\leq\Delta F(d).
\label{eq14}
\end{equation}
\noindent
Here, the minimum detectable force $\Delta F(d)$ varied between 0.1 and
0.2\,fN at different separation distances.

In Fig.~2, the constraints following from the inequality (\ref{eq14}) are
shown by the line 7, which reproduces the original line in Fig.~4 of
Ref.~\citen{66}. As is seen in Fig.~2, the new constraints are the strongest
ones over a wide interaction region extending from 40\,nm to $8\,\mu$m.
Thus, the new constraints of Ref.~\citen{66} significantly, up to a factor
of $10^3$, strengthen the results of several previous experiments, which
is a major progress in the field.

\section{Constraining Axion from the Casimir-Less Experiment}

The new Casimir-less experiment\cite{66} allows to strengthen constraints not only on
the non-Newtonian gravity, but on the axion-nucleon coupling constants as well.
The differential force due to two-axion exchange was calculated\cite{67} using
Eq.~(\ref{eq6}) with account of material and isotopic composition of the test bodies.
Note that in fact the disc used consisted of the alternating concentric strips of Au
and Si rather than of sectors, but this does not influence the values of the
additional differential force.

The constraints on $g_{an}$ and $m_a$ were found from the inequality\cite{67}
\begin{equation}
|F_{\rm add}^{\rm Au,\,Au}(d)-F_{\rm add}^{\rm Au,\,Si}(d)|
\leq\Delta F(d),
\label{eq15}
\end{equation}
\noindent
which means that no signal due to any hypothetical force was registered.
Note that if both attractive forces ($F_{\rm add}$ due to two-axion exchange
and $F_{\rm add,\,Yu}$ due to exchange of one scalar particle) exist in
nature  and contribute to the measured differential force, the constraints
imposed on each of them by the data would be even stronger than those obtained in
Refs.~\citen{66} and \citen{67} from the inequalities (\ref{eq14}) and (\ref{eq15}),
respectively.
In Fig.~1 the constraints on the axion-nucleon interaction of GUT axions following
from Eq.~(\ref{eq15}) are shown by the line 9. As is seen in Fig.~1, the constraints
of line 9 significantly improve the previously known constraints of lines 4 and 7
in the wide region of axion masses from $1.7\times 10^{-3}$ to 0.9\,eV.
The largest strengthening by a factor of 60 holds for $m_a=4.9\times 10^{-3}\,$eV.
The obtained strengthening is not as strong as in the case of Yukawa-type
corrections to Newtonian gravity. This is explained by the fact that for axions
it was necessary to use the process of two-axion exchange, whereas for gravitation
the exchange of one scalar particle was considered.  We note also that the constraints
of this and previous sections obtained from the Casimir-less experiment are free of any
problem in theoretical understanding of the Casimir force discussed in Secs.~3 and 4.

\section{Conclusions and Discussion}

In the foregoing, we have summarized the most recent results on constraining the
axion-nucleon coupling constants and the Yukawa-type corrections to Newtonian gravity
from the Casimir effect. It was shown that the major progress was achieved very recently
in both these fields and even more can be expected in near future with increasing
precision in measurements of the Casimir interaction.

It is pertinent to note, however, that all the constraints on axion-nucleon coupling
constants following from measurements of the Casimir interaction, gravitational and
Casimir-less experiments
were obtained by using the Lagrangian density (\ref{eq3})
and the effective potential (\ref{eq6}) due to an exchange of two axions.
Thus, all these constraints are valid for only the GUT axion-like
particles. There is an unresolved problem what is the effective potential due
to exchange of two QCD axions between two nucleons. If this problem were solved,
the already performed experiments on measuring the Casimir interaction could be
used for constraining the parameters of QCD axions.

As another prospective opportunity we propose an experiment on measuring the
Casimir interaction between spin-polarized (magnetized) test bodies.
In this case the simplest process of one-axion exchange between the two
nucleons and the effective potential (\ref{eq5}), valid for both the QCD
and GUT axions, can be used because it leads to a nonzero result.
The constraints obtained in such a way would be not only significantly stronger,
but much more universal by being applicable to all kinds of axions and axion-like
particles.

\section*{Acknowledgments}

The author is grateful to V.\ B.\ Bezerra, G.\ L.\ Klimchitskaya and
C.\ Romero for the collaboration on this subject.
Special gratitude is to R.\ S.\ Decca who kindly provided me with
the numerical data of the line 7 in Fig.~2.


\end{document}